\documentclass[11pt,twoside]{article}

\usepackage{asp2006}
\usepackage{epsf}
\usepackage{psfig}
\usepackage{lscape}
\usepackage{graphicx}
\usepackage{graphics}

\begin{document}
\title{Optical/UV/X-ray Insights into the RL--RQ Dichotomy}   
\author{J. Sulentic$^{1}$, S. Zamfir$^2$ and P. Marziani$^3$}   
\affil{$^1$Instituto de Astrof\'isica de Andaluc\'ia, $^2$University of Alabama, $^3$INAF, Osservatorio Astronomico di Padova}    

\begin{abstract} 
We explore the relationship between radio-loud (RL) and radio-quiet (RQ) quasars 
using a set of optical/UV/X-ray measures that are quite independent of radio measures. 
We find RL sources to show larger average FWHM H$\beta$, weaker FeII emission, no 
soft X-ray excess and no CIV blueshift--all characteristics manifested by a large 
fraction of RQ quasars (that we call Population A). We find that log L$_{1.4Ghz}$ =31.6 
ergs s$^{-1}$ Hz$^{-1}$ (or R=70) is the lower limit for RL quasars showing 
FRII morphology. We find no evidence for a hidden FRII population below this level.  
We conclude that RL sources are a distinct quasar population that may also include 
30-40\% of RQ sources which apparently show similar geometry and kinematics (what 
we call Population B). This RQ overlap, if not coincidental, may include inactive RL 
quasars as well as quasars with geometry/kinematics similar to RL sources but 
where RL activity is inhibited in some way (e.g. host morphology, BH spin).  
\end{abstract}


\section{Introduction}

Quasars were discovered because of their excess radio continuum emission yet, 
ironically, today we find only 7-10\% of them to be radio-loud (RL). The vast 
majority ($\sim$92\%) of AGN are radio-quiet (RQ) One can see from the 
broad-band spectra displayed in NED that RL sources are dominated by a 
non-thermal power-law at all wavelengths. What is the relationship between 
the RL and RQ quasars? Are they indistinguishable at other wavelengths? Both 
RQ and RL quasars, for example, show broad emission lines (unless they are 
Blazars). Are the measured properties of the lines (e.g. width, equivalent 
width, line shape and rest frame displacement) similar?  Such questions can 
be difficult to answer using the majority of available spectroscopic data 
because at low resolution and s/n all quasars look alike. The advent of 
the Sloan Digital Sky Survey offers us more uniform, high resolution and 
high s/n data. This is  especially true if one restricts oneself to the 
brightest (g$\leq$17) SDSS quasars.

In order to test the existence of a RL-RQ dichotomy we require an observational 
context within which to search for differences. This context should involve 
observational parameters that are  independent of radio measures. We have been 
developing a four dimensional parameter space that builds upon pioneering 
studies of the PG quasars (Boroson \& Green 1992; Wang et al. 1996). This attempt at 
a multiwavelength  unification of quasar properties (4DE1) involves optical, UV and 
X-ray measures (Sulentic et al. 2000, 2007; Marziani et al. 2001, 2003a). Our four 
principal parameters are: 1) full width at half maximum of the broad H$\beta$ 
emission line (FWHM H$\beta$), 2) the equivalent width ratio of FeII4570 
blend$\over$H$\beta$ (RFE), 3) the FWHM  normalized centroid shift of CIV1549 
(C(1/2)) and 4) the soft X-ray photon index ($\Gamma_{soft}$)  (Marziani et al. 
2001, 2003a). We have addressed the RL-RQ question using both our own atlas 
sample (Marziani et al. 2003b) and a bright SDSS DR5 sample. Both are low z samples 
largely brighter than g=17.5 (Zamfir et al. 2008).

\begin{figure}
\includegraphics[scale=1.04, bb = -20 0 250 215]{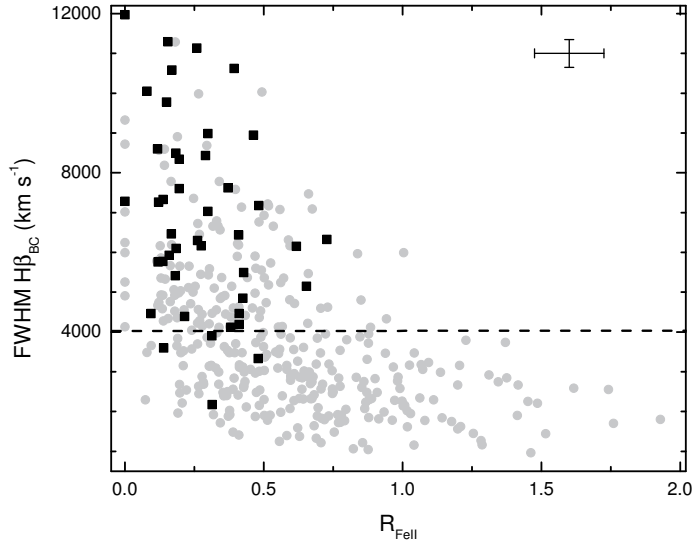}
\caption{Optical plane of the 4DE1 parameter space showing the
distribution of the SDSS quasars FWHM of H$\beta$ versus the
FeII-prominence parameter RFE. Filled squares identify FRII sources, while
the grey spot represent the general quasar population.}
\end{figure}
\section{Optical Eigenvector Plane}

Figure 1 shows the distribution of the 400+ brightest (g$\leq$17.0) SDSS DR5 
(z$\leq$0.7) quasars in the optical plane of 4DE1. Sources with the broadest 
Balmer lines fall at the top with narrow-line Seyfert 1 (NLSy1) sources at 
the bottom. Sources with the strongest FeII emission lie toward the right.
The distribution of sources is clearly non-random with sources showing both broad 
Balmer lines and strong FeII emission being absent. The SDSS sample shows that 
FeII emission is almost ubiquitous in broad line AGN. The detection of weak 
FeII in many of the broadest sources indicates that the zone of avoidance in the 
upper right of the diagram is real. The trend from weak FeII/very broad H$\beta$ 
to strong FeII/narrow H$\beta$ is no doubt driven by a combination of broad line 
region geometry/kinematics convolved with source orientation. The horizontal lines 
corresponding to FWHM H$\beta$=4000km/s marks the separation between what we call 
Populations A and B quasars. 
Sources above and below this line show many
differences (Sulentic et al. 2007). In the 4DE1 context
sources below the line show a blueshifted/asymmetric
CIV1549 profile as well as a soft X-ray excess. Narrow line
Seyfert 1 (NLSy1) sources are found there. Sources above
the line show a more symmetric  unshifted CIV profile
and no soft X-ray excess.  So the 4DE1 parameters 
show considerable diversity among the brightest 400+ SDSS quasars. The relevant question 
for this report is whether RL sources occupy the same parameter domain as the RQ majority.

\begin{figure}
\includegraphics[scale=1.00, bb = -20 0 250 215]{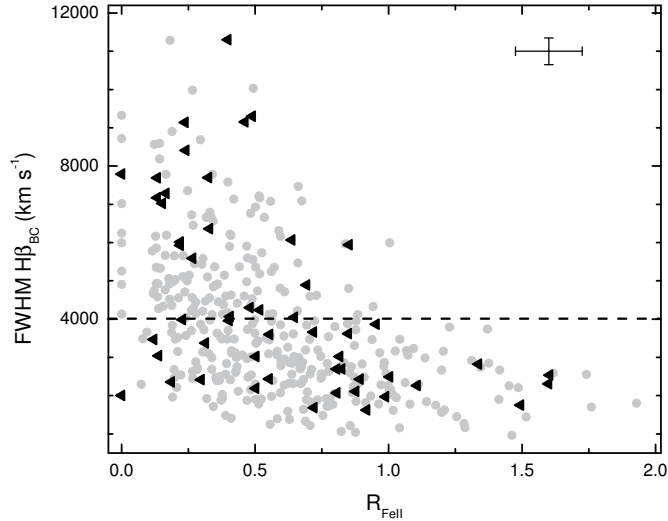}
\caption{Same as for the previous Figure, but with filled triangles now
indicating intermediate radio sources.}
\end{figure}

Sources (n=48) showing edge-brightened double-lobed radio (FRII) structure on FIRST 
(supplemented by NVSS) maps are indicated with black squares to distinguish them 
from the RQ majority 
(see Fanaroff \& Riley 1974). This is the most unambiguous RL source population. Sources 
showing FRII radio morphology also show radio/optical flux ratios R$\geq$70 (e.g. 5Ghz and 
4100 $\AA$; Kellermann et al. 1989) and  log L$_{1.4Ghz}$$\geq$31.6 ergs s$^{-1}$ Hz$^{-1}$.
Below these values all sources show weak core and/or core-jet radio morphologies. It is 
visually apparent that RL quasars do not occupy the full parameter domain defined by RQ 
sources but a restricted domain largely above FWHM H$\beta$=4000km/s and and below RFE=0.5. 
RL sources are Population B quasars using our designation. A 2D K-S test confirms at a high 
level of confidence that the RQ and RL  populations do not occupy the same domain (Zamfir et 
al. 2008). This is clear evidence that RL and RQ sources are fundamentally 
different in some structural and/or kinematic way. We note that as many as 40\% of RQ 
sources also occupy the domain defined by FRII RL sources.  Either the RL-RQ or Pop. A-B 
distinction involves something fundamental. We have recently argued that the Pop. A-B 
differences are stronger than the RL-RQ ones. Either way, FWHM$\approx$4000km/s emerges 
as a boundary possibly corresponding to a critical Eddington ratio$\approx$0.2$\pm$0.1 
where BLR structure and kinematics apparently undergo a significant change (Sulentic et al. 2007).

So the RL-RQ dichotomy is only partial when considered in a 4DE1 context. The log L$_{1.4Ghz}$
(or R) cutoff for sources showing FRII morphology is rather sudden. Could it be more diffuse with 
many less radio luminous sources showing FRII structure too weak to be detected by 
FIRST? Comparison of FIRST and NVSS fluxes for the FRII sample always reveals an NVSS excess reflecting 
FIRST insensitivity to much of the extended FRII structure. However NVSS and FIRST fluxes are almost 
always the same for CD sources near log L$_{1.4Ghz}$$\geq$31.6 ergs s$^{-1}$ Hz$^{-1}$. There is therefore 
no evidence for a significant number of hidden FRII sources. Perhaps weaker FRII sources are associated 
with fainter quasars and were not included in our bright quasar sample? Fortunately we can 
appeal to a search for FRII RL emission from SDSS DR3 quasars down to g$\approx$19 (de Vries et al. 2006) 
which provides n=67 additional sources. None cross our previously defined radio and bolometric luminosity
boundaries. If the RL-RQ dichotomy is real then we expect a sharp boundary reflecting a discontinuity 
between radio emission from RL and RQ sources. 

If there were a continuous distribution of radio properties among all quasars then we might find the 
bridge or transition objects at intermediate radio (RI) luminosities (see e.g. Falcke et al. 1996). In 
Figure 2 we consider the range log L$_{1.4Ghz}$=31.0-31.6 ergs s$^{-1}$ and ask if they distribute 
like the bona-fide 
FRII RL sample. Figure 2 shows that they do not--but instead RI distribute like RQ quasars. We conclude 
that there is a discontinuity in the radio properties between RL and RQ sources at log L$_{1.4Ghz}$=31.6. 
RI are marked as black squares and RQ as grey dots in Figure 2.
This does not preclude a RQ quasar from showing weak core-jet structure--galactic sources can show a core 
jet structure--but this does not make them RL AGN.  And weak lobes have been detected from a few Seyfert 
galaxies (e.g. NGC3367, R$\sim$1.0; Garcia-Barreto et al. 2002). Deeper radio searches have not revealed 
weak lobes in RI quasars (e.g. Lu et al. 2007). No evidence exists to suggest that RI sources bridge the RL-RQ dichotomy.

\begin{figure}
\includegraphics[scale=1.02, bb = -30 0 250 215]{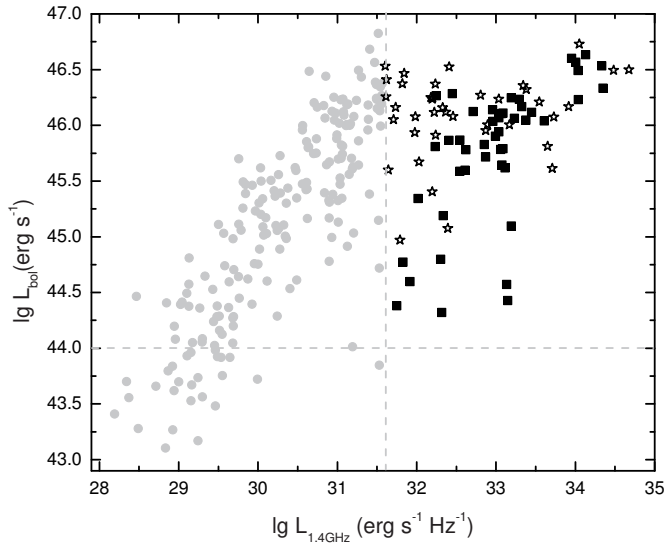}
\caption{Bolometric luminosity vs. specific radio luminosity at 1.4 GHz
for the quasar of the SDSS sample.  Filled circles represent FRII sources;
stars, core dominated sources.}
\end{figure}

\section{Too Many CD RL Sources?}

Note that we have omitted core-dominated (CD) RL sources in Figure 1. They would be considered as RL using 
our radio luminosity and R definitions with lower boundary set by the weakest FRII 
sources. We omitted them because their nature is more ambiguous. Figure 3 plots 
FRII (black squares) and CD RL (stars) as well as all other radio detected RQ quasars (grey squares) 
in log L$_{BOL}$ vs  log L$_{1.4Ghz}$ space. We see a clear RQ trend for our bright SDSS  
sample whose lower edge is set by the radio and optical flux limits of our sample. We also see a well defined
FRII RL trend above log L$_{1.4Ghz}$=31.6 and above log L$_{BOL}$=44.0. The region between the parallel 
trends would be filled by radio-detected RQ quasars in deeper surveys (zamfir et al. 2008).
In standard orientation-unification scenarios FRII sources are the parent population of RL quasars. 
CD RL sources are interpreted as FRII sources with jet axis aligned near our line of sight. A 
too close alignment yields a blazar (like BLLAC) where the continuum swamps the broad optical lines 
(obviously excluded from our sample). Apparently many less well-aligned sources produce a CD source with 
a boosted power-law continuum but not boosted enough to swamp the broad lines. These involve CD sources 
in Figure 3 that fall rightward of the FRII sources.  We see perhaps 15 such (assumed relativistically boosted) 
CD source in Figure 3. 

The remaining CD RL sources make no sense in an orientation-unification scenario. An aligned FRII cannot 
be radio fainter than a mis-aligned one. The 20+ CD RL that fall toward the radio weak side of the RL 
domain must be either: 1) boosted sources from an unseen FRII population even further to the left in 
Figure 3, 2) the most radio-bright RQ quasars, possibly even modestly boosted or 3) ``birthing'' 
RL sources. We have already argued that there is no current evidence for a population of FRII sources 
below log L$_{1.4Ghz}$=31.6 ergs s$^{-1}$. Option 2 is favored because many of the CD sources in question 
continue the well defined radio-bolometric luminosity trend for RQ quasars see in Figure 3. This means that 
between log L$_{1.4Ghz}$$\approx$31.6-32.6 ergs s$^{-1}$ there is RQ overlap with FRII sources in the 
FRII RL parent population. Option 3 has been considered in connection with ``Ghz-peaker'' sources (O'Dea 1998).
Unfortunately most of these sources (including RI) lack multi-frequency radio measures.

In connection with option 2 above we note that 16 CD ``RL'' sources would fall below FWHM H$\beta$=4000km/s 
in Figure 1 where only 4 FRII sources are found in a domain where 98\% of the sources are RQ. Thus their 
apparent radio loudness is especially suspicious. 15/16 of these sources fall leftward of the FRII 
sources in Figure 3 leading us to conclude with more 
confidence that they are RQ sources with above average radio luminosity and not classical RL quasars. 
If we assume that the $\approx$25 CD sources on the left side of the FRII population 
are RQ quasars we reduce the number of bona fide RL sources by $\approx$25\% and further reduce the 
RL quasar fraction from 5-9\% (Zamfir et al. 2008) to 6\%$\pm$2\%. This would reduce the inferred 
probability of radio loudness in Population A to below 1\%.

Including both FRII and CD RL sources in Figure 1 allows one to assess the role of source orientation
in the optical 4DE1 plane. FRII and CD sources show quite different mean FWHM H$\beta$ and RFE values with 
a 2D KS test giving a probability P$\approx$10$^{-4}$ that they show the same domain occupation.
CD RL sources concentrate around FWHM H$\beta$$\approx$4000km/s and RFE$\approx$0.5 while FRII RL 
concentrate around FWHM H$\beta$$\approx$6700km/s and RFE$\approx$0.3. Outlier CD and FRII sources with 
very high and very low FWHM H$\beta$ values respectively can be viewed as misaligned sources.  This work 
confirms with a more complete sample earlier attempts to find an orientation sequence related to
FWHM H$\beta$ (Wills \& Browne 1986, Rokaki et al. 2003, Sulentic et al. 2003). 
  
RL sources show differences from a large fraction of RQ sources. The differences extend to many parameters 
beyond the 4DE1 measures mentioned here. These differences and the robust lower radio luminosity
boundary for FRII morphology (assumed to be the RL parent population), argues that RL are not simply the bright 
end of the radio luminosity distribution for all quasars. They also argue that BLR properties for RL sources are 
distinctly different from BLR properties for a large fraction of RQ quasars we call them Population A). The main 
question involves the 30-40\% of RQ sources that occupy the same domain as RL quasars (population B). This overlap 
region may simply be coincidental--reflecting the large parameter domain occupied by RQ sources. Our previous work  
suggests that RL and RQ pop B. sources show larger BH masses and--especially--lower Eddington ratios. If the RQ 
overlap population belong to the same physical regime as RL sources then there may be two kinds: 1) currently 
inactive RL quasars and 2) quasars with BLR properties similar to those of RL sources but where one or more 
additional properties inhibit the onset of FRII RL activity (e.g. host galaxy morphology and BH spin).  

\acknowledgements We thank the organizing committee for the generous allocation of time. JWS acknowledges 
support from Junta de Andalucia P08-FQM-4205-PEX.

\vfill\vfill

\begin{thebibliography}{}
\bibitem[Boroson \& Green(1992)]{} Boroson, T. \& Green, R. 1992, \apj, 80, 109 
\bibitem[de Vries et al.(2006)]{} de Vries, W. et al. 2006, \aj, 131, 666 
\bibitem[Falcke et al.(1996)]{} Falcke, H. et al. 1996, \apj, 471, 106 
\bibitem[Fanaroff \& Riley(1974)]{} Fanaroff, B. \& Riley, J. 1974, \mnras, 167, 31 
\bibitem[Garcia-Barreto, et al.(2002)]{} Garcia-Barreto, J. et al. 2002, \aj, 123, 1913 
\bibitem[Kellermann et al.(1989)]{} Kellermann, K. et al. 1989, \aj, 98, 1195 
\bibitem[Lu et al.(2007)]{} Lu, Y. et al. 2007, \aj, 133, 1615 
\bibitem[Marziani et al.(2001)]{} Marziani, P. et al. 2001, \apj, 558, 553
\bibitem[Marziani et al.(2003a)]{} Marziani, P. et al. 2003a, \mnras, 345, 1133 
\bibitem[Marziani et al.(2003b)]{} Marziani, P. et al. 2003b, \apjs, 145, 199  
\bibitem[O'Dea et al.(1998)]{} O'Dea, C. et al. 1998, \pasp, 110, 493 
\bibitem[Rokaki et al.(2003)]{} Rokaki, E. et al. 2003, \mnras, 340, 1298 
\bibitem[Sulentic et al.(2000)]{} Sulentic, J. et al. 2000, \apj, 536, L5
\bibitem[Sulentic et al.(2003)]{} Sulentic, J. et al. 2003, \apj, 597, L17 
\bibitem[Sulentic et al.(2007)]{} Sulentic, J. et al. 2007, \apj, 666, 757 
\bibitem[Wang, et al. (1996)]{} Wang, T. et al. 1996, \aap, 309, 81 
\bibitem[Wills et al.(1986)]{} Wills, B. et al. 1986, \apj, 302, 56
\bibitem[Zamfir et al.(2008)]{} Zamfir, S. et al. 2008, \mnras, 387, 856 
\end{thebibliography}
\end{document}